\documentclass[12pt]{elsart}
\usepackage{natbib}
\usepackage[latin1]{inputenc}
\usepackage{amssymb}
\usepackage{graphics}
\begin{document}
\begin{frontmatter}
\title{ Bifurcations and Averages in the Homoclinic Chaos 
of a Laser with a Saturable Absorber}
\author{Hugo L. D. de S. Cavalcante}, 
\author{Jos\'e. R. Rios Leite\corauthref{rios}}
\ead{rios@df.ufpe.br}
\corauth[rios]{Correspondig author. Tel.: +55-81-32718450; fax:+55-8132710359}
\date{1 August 2000}
\address{Departamento de F\'{\i}sica,  Universidade Federal de
Pernambuco, 
50670-901 Recife, PE Brazil  }
\begin{abstract}
The dynamical bifurcations of a
laser with a saturable absorber were calculated,
with the 3-2 level model, as function of the  gain parameter.
The average power of the laser is shown to have specific
behavior at bifurcations. The  succession of periodic-chaotic windows,
known to occur in the homoclinic chaos, was studied numerically.
 A critical exponent of 1/2
is found on the tangent bifurcations from chaotic into periodic
pulsations.
\end{abstract}

\end{frontmatter}

\section{Introduction}

Chaos in monomode $CO_{2}$
lasers with an intracavity molecular gas as saturable absorber, $LSA$,
has been observed
\cite{Dangoisse88,Hennequin88,Tachikawa88,Zambon91,Papoff91,Lefranc91}
and modeled by a three level system to represent the amplifier and a
two level system for the
absorber \cite{Tachikawa88}.
A passive Q-switching of the laser cavity is produced by the saturation of
the absorber
and, for appropriate relaxation and saturation rates,
the laser oscillates with  periodic or chaotic pulse emission.
The dynamical system behavior is
explained as due to the approach to a
homoclinic tangency to a saddle focus \cite{Hennequin88}, 
or to an unstable periodic orbit
 \cite{Zambon91,Papoff91,Lefranc91},
depending on the operating parameters of the system.

This work presents numerical solutions of the laser model giving the
dynamical bifurcations as it
approaches a homoclinic orbit to an unstable cycle
which consists of a high spike pulse and an
infinite number of undulations.
Bifurcations diagrams are usually exhibited
by giving the peak of the pulses and here they are given by
the average power of the laser.
The properties of the average power throughout the bifurcations are the
main results reported.

 \section{The 3-2 level model}

Introduced by Tachikawa {\it et al} \cite{Tachikawa88}
and thoroughly  analyzed by many authors
the $3-2$ level model for the $LSA$ predicts most
dynamical observations made in these lasers.
The mean intracavity intensity $I(t)$,
the mean gain medium population inversion $U(t)$,
the effective ground state population depletion of
the gain medium, $W(t)$, and
the mean population
difference of the absorber $\overline{U}(t)$,
are the four independent variables
for the rate equations in the model.
It can be shown \cite{Lefranc91,deOliveira97} that the LSA system
may be reduced to a three dimensional flux as one
considers the usual fast relaxation of  the absorber \cite{deOliveira97}, and
adiabatically eliminates the
variable $\overline{U}(t)$.

 The three level-two level rate equations describing the $LSA$
with the normalized variables explained above are \cite{Lefranc91}:
\begin{eqnarray}
\dot{I} & = & I \left( U - \overline{U}-1 \right) \\
\dot{U} & = & \epsilon \; \left[ W - U \left( 1 + I \right) \right] \\
\dot{W} & = & \epsilon \; \left( A + b U - W \right) \\
\dot{\overline{U}} & = & \overline{\epsilon} \; \left[ \overline{A} -
\overline{U} \left( 1 + a I \right) \right]
\end{eqnarray}
The parameters in the equations are:  $\epsilon$ and $\overline{\epsilon}$
the relaxation rates of
the amplifier and absorber, respectively,
normalized to the cavity relaxation rate (which is the inverse time unit).
$A$ and $\overline{A}$  are the pumping rates in the
amplifier and absorber population differences, respectively,
normalized to the cavity loss.
The coefficient $b$ is the difference between the population relaxation rates
of lower and upper level of the gain transition,  normalized
to the sum of these relaxation rates.
The saturability coefficient for the absorber, normalized to
the two level saturability of the gain, is $a$.
More detailed explanation of these dimensionless quantities and the
physical behavior contained in Eqs. (1)-(4) are given in Ref. \cite{Lefranc91}.

The flux described by those
equations was studied by Lefranc {\it et al.} \cite{Lefranc91} who
calculated bifurcations diagrams typical of homoclinic chaos.
Generally each laser pulse  has a leading spike, named the reinjection,
and $ (0,1, \ldots ,n, \ldots)$
small undulations in a tail associated to the cycling around the unstable
orbit to be reached at the homoclinic tangency.
The notation is then $P^{(n)}$ for the pulses with $n$ undulations
and the periodic regime made with such pulses and $C^{(n)}$ for the chaotic
pulsation composed of irregular occurrence of pulses $P^{(m)}$ with 
$m\leq n+1$.
The maxima in the spike and the undulations of a long train of  pulses
can be registered to indicate the dynamical state of the LSA.
A bifurcation diagram
consists of a plot of the maxima in the pulses emitted versus the
control parameter $A$,
which is the pump rate of  the laser.

The bifurcations, as shown in Figure $1(a)$, consist of alternating
periodic-chaotic windows
well predicted \cite{Lefranc91} in the homoclinic chaos of LSA.
At certain  chosen parameters they also show  long sequences windows 
separated by periodic to periodic global bifurcations 
from a
regime $P^{(n)}$  to a $P^{(n+1)}$ \cite{Hugopre00}.

\begin{figure}[htbp]
\begin{center}
\rotatebox{-90}{\resizebox{9cm}{!}{\includegraphics{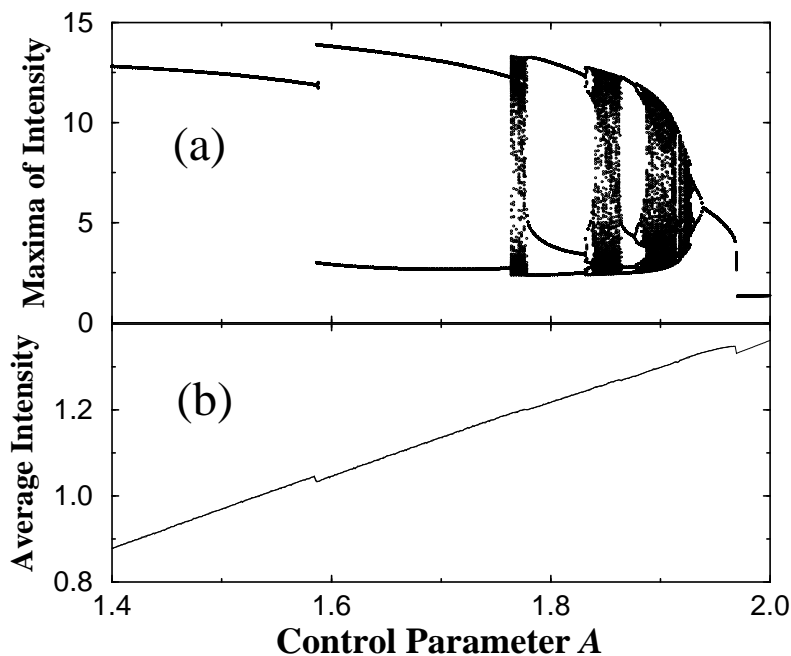}}}
\end{center}
\caption
{   Bifurcation diagram calculated
using the 3-2 level
model for the gain and the absorber, respectively.
The parameter choice was made following ref \cite{Lefranc91}~:
(a) gives the peak value of the pulses and (b) the corresponding
average power. Notice  significant
variation of the average power at the bifurcations between periodic regimes
$P^{(2)}$ to $P^{(3)}$ and at the Hopf bifurcations, on the end of the diagram.
At the period doubling and at the  tangent
bifurcations the changes in the average are not visible. }
\end{figure}

 \section{Averages and Bifurcations}

The average of a dynamical variable in a nonlinear flux or map
may manifest some of its bifurcations \cite{Luiz96,Hugoave00}.
Such property has been pointed out in the early days of
studies of chaotic maps \cite{Lorenz63}. It is our purpose here
to verify how such average behaves in a flux associated to a laser system where
experimental tests are feasible \cite{Luiz96,Hugoavex00}.

The parameters used to numerically solve  equations (1-4)
were: $\overline{A} = 2.16$,
 $\epsilon = 0.137$, $\overline{\epsilon} = 1.2$, $a = 4.17$
and
 $b = 0.85$. The control parameter $A$ was varied between $1.4$ and $2.0$.
The bifurcation diagram showing the maxima of the pulses
is given  in Figure $1(a)$. Lefranc {\it et al} \cite{Lefranc91} have
discussed this
diagram in details. At each position of $A$ $400$ maxima are saved
from the numerical integration done with a standard fourth order
Runge-Kutta routine. The calculation at each of the $300$ steps of $A$ had
as
initial condition the last values from the previous step.
For low gain, i.e. small $A$, the laser shows
periodic pulsation with spike pulses $P^{(0)}$ which bifurcates into
periodic pulsation of one undulation pulses. This global
bifurcation consists in a transition that generally has bistability.
Increasing the value of $A$ gives a cusp bifurcation from
periodic $P^{(2)}$ into chaotic $C^{(2)}$ pulsation. The chaotic window ends
at a tangent bifurcation, when the system recovers periodic
pulsation $P^{(3)}$. This  $P^{(3)}$  window develops a cascade of
period doubling bifurcations into $C^{(4)}$ chaos and so on.
The succession of windows is one of the signatures of the homoclinic chaos.
At $A =1.92$ an inverse cascade of period doubling leads the system to a Hopf
bifurcation connected to the continuous wave laser operation, when the fixed point
$I_{+} \neq 0$ is stable.
Figure $1(b)$ shows the average power of the laser calculated in same range
of variation of the control parameter $A$. The transient at each step of the
calculation, which is very important near the bifurcation points, was
eliminated, as far as possible within computation time, by neglecting the
first $25000$ time units in a total of
$80000$ time units of integration.
It can be seen that the most important manifestation of the bifurcations
in the average occur at the periodic to periodic transition and at
the Hopf bifurcation.
 The average suffers a discontinuity at the periodic to periodic
transitions. Such discontinuity also occurs at the Hopf bifurcation, when
the parameters make this a supercritical one. At the period doubling and
at the tangent
 bifurcations the changes in the average are not visible in the scale
resolution of Figure $(1b)$. To search for the
average variations on these bifurcations a refined calculation was done over
 the period doubling and the tangent bifurcation around the $C^{(3)}$
chaotic window, as shown in Figure $2$.
The period doubling bifurcation remains without changes for the average.
At these bifurcations the only expected changes on the average is  in one 
of its derivative with respect to the control parameter. 
For the flux studies  here no derivative discontinuity was obtained to the
resolution of the calculation.
 Conversely a first derivative discontinuity is obtained, analytically and
numerically,  for the average of  one dimensional maps at period doubling
bifurcations \cite{Hugoave00}.   
\begin{figure}[tbp]
\begin{center}
\resizebox{9cm}{!}{\includegraphics{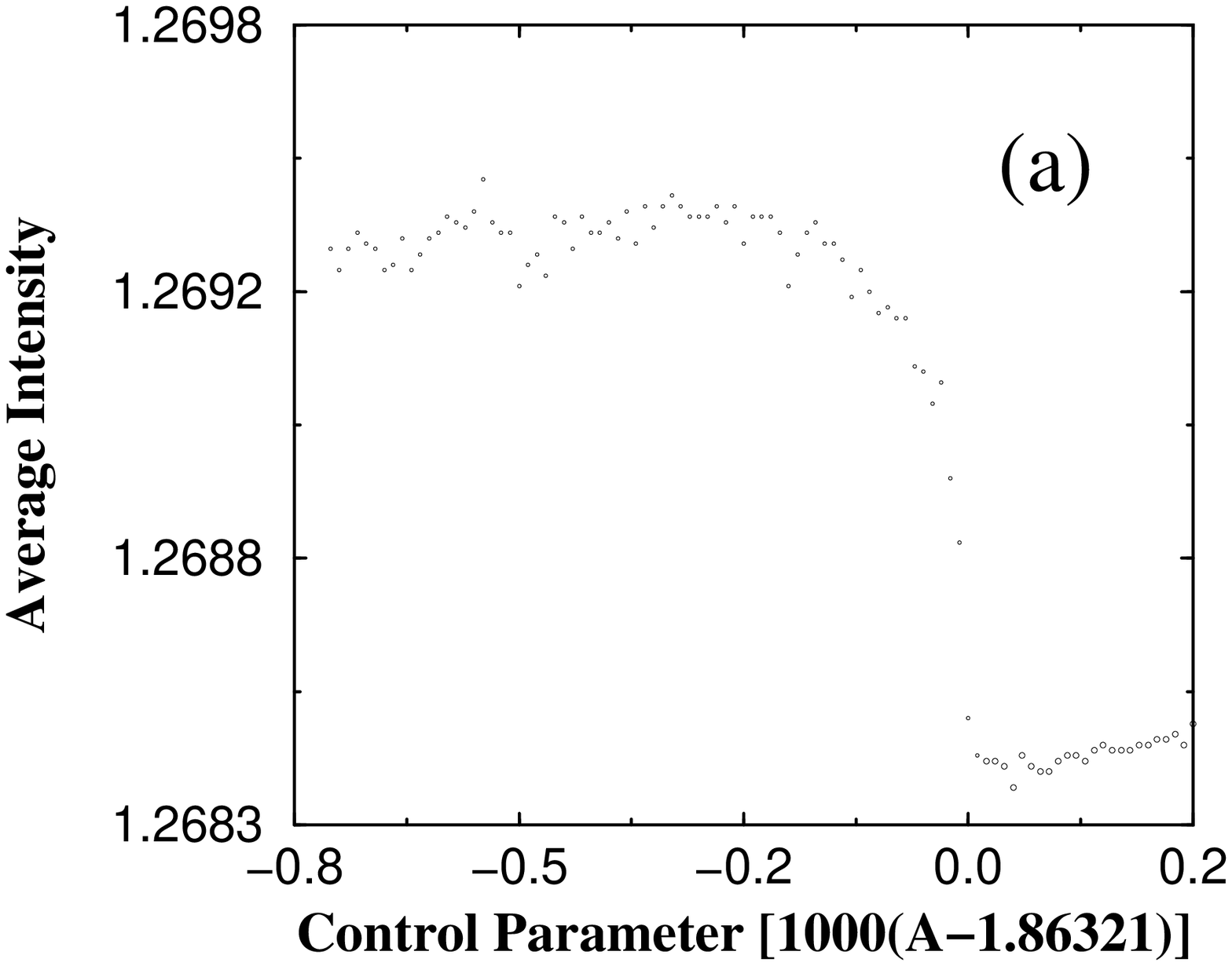}}
\end{center}

\begin{center}
\rotatebox{-90}{\resizebox{6cm}{!}{\includegraphics{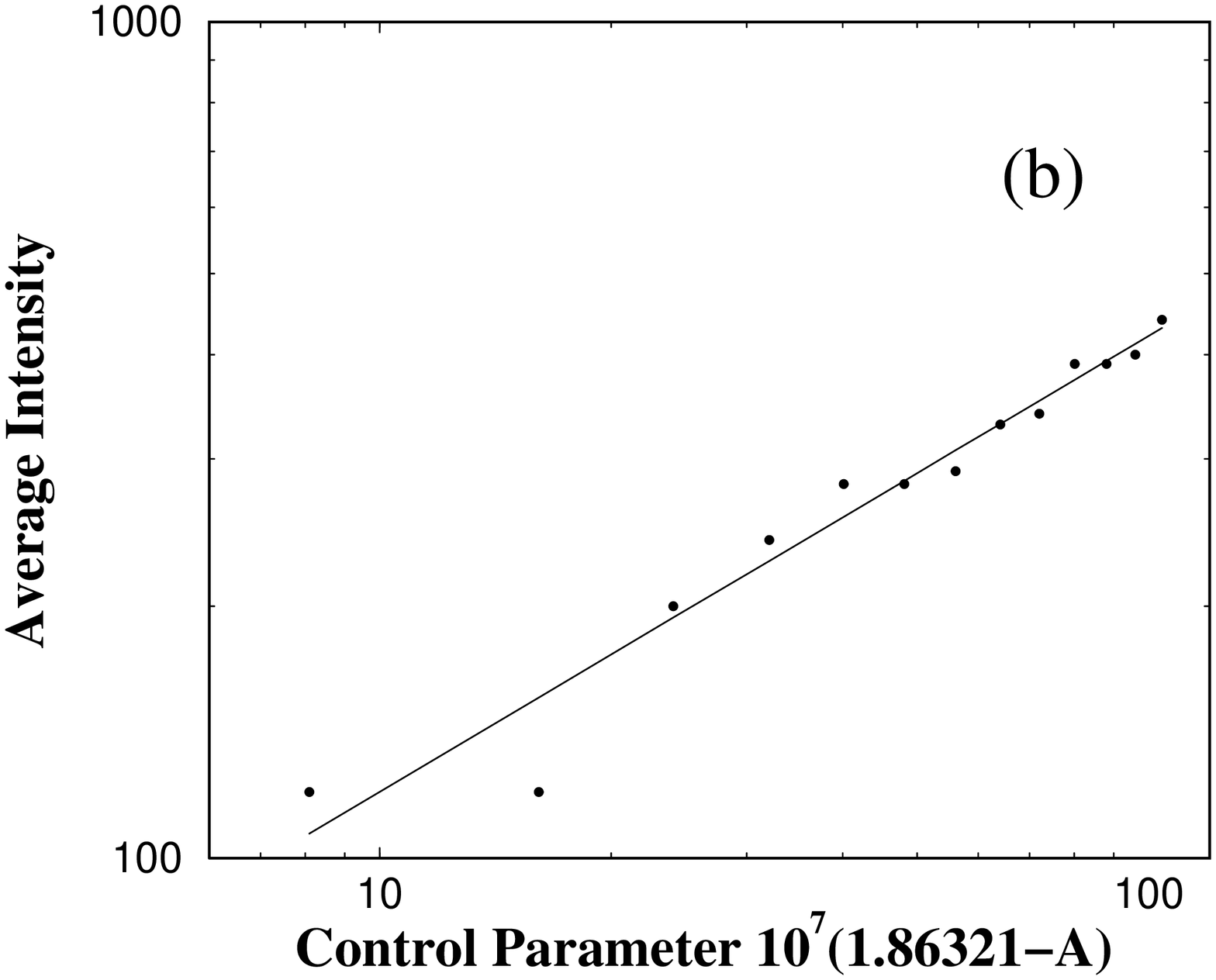}}}
\end{center}
\caption
 {Bifurcation diagram for the average laser power obtained numerically
 using the 3-2 level
 model: $(a)$ shows a detail of Figure 1 with the  tangent bifurcation
 from chaotic regime $C^{(3)}$ to periodic $P^{(4)}$.$(b)$ gives 
a log-log graph where the average decreases following a power law that is 
fitted to the exponent $\nu = 1/2$.}
 \end{figure}
Figure $2$ shows a bifurcation diagram  for the average
still with the same
parameters of Figure $1$, but spanning  a very narrower range of $A$ 
at the onset
of the tangent bifurcation
 from chaotic regime $C^{(3)}$ to periodic $P^{(4)}$. Similar
behavior
occur for the other tangent bifurcations.
To fit the  dependence of the average ${\overline I}$ on $A$ 
the expression
\begin{equation}
{\overline I}(A) = {\overline I}_{i} +
B_{i}(A^{c}_{i}-A)^{\nu_{i}}
\end{equation}
was used, with the values of ${\overline I}_{i}$ and $A^{c}_{i}$ extracted from
inspection of the
numerical diagram near the $(i)$ bifurcation . 
A best fitting of eq.~(5) to the data is shown in Figure $(2b)$ in a 
log-log graph.
The dots are the numerical calculation and the line corresponds to the fitting 
which verifies that the  average decreases following a power law whose  
exponent is $\nu_{i} = 1/2$. The other tangent bifurcation in the diagram 
seems to
have the same exponent but a detailed study was not pursued.

\section{Conclusions}
Bifurcation diagrams associated  to the homoclinic tangency to  an
unstable periodic orbit in the 3-2 level model for a laser with 
saturable absorber
were calculated numerically.

The numerical solutions were  used to calculate
the average power emitted by the chaotic laser.
It was verified that the average power
has a discontinuous change on the cusp bifurcations, related to the
bistability between different
periodic regimes.
Through period doubling bifurcations, no change of the average power 
was obtained.
 The tangent bifurcations from chaos
into periodic pulsation show a characteristic
average power dependence, similar to the averages observed  on
the logistic map \cite{Hugoave00}, with a critical exponent 1/2.
Preliminary experiments
observing bifurcations in the average power of a $CO_{2}$ laser with $SF_{6}$ 
as saturable
absorber have been done \cite{Luiz96} and the verification of the
properties of critical exponents on the tangent bifurcations is under
development \cite{Hugoavex00}.
Characterizing Dynamical bifurcations by averages, which can be measured
experimentally by slow detectors, may be useful to study ultra fast chaotic
systems.
\section*{Acknowledgments} Work partially supported by Brazilian
Agencies: Conselho
Nacional de Pesquisa e Desenvolvimento (CNPq), 
Financiadora de Estudos e 
Projetos (FINEP) and part of the Pronex project on Nonlinear Optics, Lasers 
and Applications


\begin{thebibliography}{9}

\bibitem{Dangoisse88} 
D. Dangoisse, A. Bekkali, F. Papoff  and P. Glorieux,
 Europhysics~Lett.~{\bf6} (1988) 335.

\bibitem{Hennequin88} 
D. Hennequin, F. de Tomasi, B. Zambon and E. Arimondo,
Phys. Rev.~A{\bf 37} (1988) 2243.

\bibitem{Tachikawa88} 
M. Tachikawa, F.-L. Hong, K. Tanii and T. Shimizu,
Phys. Rev. Lett. {\bf60} (1988) 2266.

\bibitem{Zambon91}  
B. Zambon,
Phys. Rev. A{\bf 44} (1991) 688.

\bibitem{Papoff91} 
F. Papoff, A. Fioretti and E. Arimondo,
Phys.Rev. A{\bf 44} (1991) 4639.

\bibitem{Lefranc91} 
M. Lefranc, D. Hennequin and D. Dangoisse,
J. Opt. Soc. Am. B{\bf 8} (1991) 239.

\bibitem{deOliveira97}
P. C. de Oliveira, M. B. Danailov, Y. Liu and J. R. Rios Leite,
Phys. Rev. A{\bf 55} (1997) 2463.

\bibitem{Luiz96}
L. de B. Oliveira-Neto, J. F. T. da Silva, A. Z. Khoury and J. R. Rios Leite,
Phys. Rev. A{\bf 54} (1996) 3405.
\bibitem{Hugopre00} Hugo L. D. de S. Cavalcante  and  J. R. Rios Leite, 
unpublished
(submitted to Phys. Rev. E, (1999))

\bibitem{Hugoave00} Hugo L. D. de S. Cavalcante  and  J. R. Rios Leite, 
Dyn. Stab.Sys. 1 (2000) 15.

\bibitem{Lorenz63} E. N. Lorenz, Tellus XVI (1964) 1.

\bibitem{Hugoavex00} Hugo L. D. de S. Cavalcante  and  J. R. Rios Leite,
unpublished.

\end{thebibliography}
 \end{document}